\documentclass[manuscript]{aastex}

\shorttitle{Planet formation}
\shortauthors{Bodenheimer \& Lissauer}

\begin{document}


\title{Accretion and Evolution of $\sim 2.5$ M$_\oplus$ Planets with 
Voluminous  H/He    Envelopes}


\author{Peter Bodenheimer}
\affil{UCO/Lick Observatory, Department of Astronomy and Astrophysics, University of California,
    Santa Cruz, CA 95064}
\email{peter@ucolick.org}

\author{Jack J. Lissauer}
\affil{Space Science and Astrobiology Division, NASA-Ames Research Center, 
Moffett Field, CA 94035}
\email{Jack.J.Lissauer@nasa.gov}

\begin{abstract}

Formation of  planets in the Neptune size range 
with low-mass, but voluminous,  H$_2$/He gaseous envelopes is
 modeled   by detailed numerical simulations
according to the core-nucleated accretion scenario. Formation locations
ranging from 0.5 to 4 AU from a star of 1 M$_\odot$ are considered. 
The final planets have
heavy-element cores of  2.2-- 2.5  M$_\oplus$ and envelopes 
in the    range 0.037--0.16  M$_\oplus$.
After the formation process, which lasts 2 Myr or less, the planets evolve 
at constant mass up to an age of several Gyr. For assumed equilibrium 
temperatures of 250, 500, and 1000 K,   their 
calculated final radii are compared with those observed by the \textit{Kepler}
spacecraft. For the particular case of Kepler-11 f, we address the question
whether it could have formed {\it in situ} or whether migration from a
formation location farther out in the disk is required.
\end{abstract}

\keywords{planets and satellites: formation--planets and satellites:
physical evolution--planets and satellites: individual (Kepler-11 f)}

\section{Introduction}
\label{sect:intro}

The analysis of the first 22 months of {\it Kepler} data shows over 2700
planetary candidates detected through transit observations \citep{bat13,bur14},
most of which orbit within 0.5 AU of their star. 
Many of these planets have radii of     2--6 R$_\oplus$, which is roughly the
Neptune size range, but most seem to be sub-Neptune in terms of mass.
A small fraction of the {\it Kepler} planets also
have mass determinations. For example, transit timing variations in the
Kepler-11 system yield masses between 1.9 and 8.0 M$_\oplus$
for  radii between 1.8 and 4.2 R$_\oplus$  \citep{lis13}.  A radial velocity survey \citep{mar14}
provides additional masses.  
 \citet{wu13} analyzed    about 30 planets,
combining their statistical data from transit timing variations with radial
velocity data from other sources to estimate a mass-radius relation in the  mass
range 2--25 M$_\oplus$.  They find
 $M \sim 3$ M$_\oplus (R/$R$_\oplus)$ with
considerable scatter, in constrast to the more standard relation often
used to derive masses from transit radii: $M/$M$_\oplus 
 \approx (R/$R$_\oplus)^2$, based on planets in the solar system  of Saturn
mass or less \citep{lrf11}. Again, many of the Wu \& Lithwick objects  fall 
into the ``sub-Neptune" class. \citet{wm13} find a mass-radius fit for
planets with radii between 1.5 and  4 R$_\oplus$, using data largely independent of
those of \citet{wu13}: $M/$M$_\oplus \approx  2.69 (R/$R$_\oplus)^{0.93}$.
\citet{wei13} include the stellar flux at the planet ($F$) in the mass-radius
relation and find a somewhat different relation, for planet masses less
than 150 M$_\oplus$: $R=1.78 M^{0.53} F^{-0.03}$, where $F$ is in units of 
erg s$^{-1}$ cm$^{-2}$ and radii and masses are in Earth units.
A list of all explanets with measured masses
and radii (as of early 2013) is found in \citet{wei13}; a list of such planets
with radii less than 4 R$_\oplus$ (as of early 2014) is in \citet{wm13}, and
one for planets of less than 100 M$_\oplus$ (as of late 2013) appears in \citet{lop13}. 
 
Given the density, from the mass and radius measurements, planets  with masses in the 
range 1--10 M$_\oplus$ can be separated
into three groups.  The higher-density group, with mean density $\bar \rho >5.0$ 
g cm$^{-3}$, must  have a ``rocky" composition  almost entirely of heavy elements. 
 The  low-density planets  ($\bar \rho <1.5$ g cm$^{-3}$)  also have a substantial
fraction of their mass in a heavy-element core of rock and (possibly) ice, but must also
have a significant fraction of their volumes occupied by light gasses. 
Intermediate-density planets can either have their volumes containing
substantial amounts of rock and of H/He,  or can be composed 
 mostly  of water and/or other
astrophysical ices. This paper focusses on the low-density group.

Model calculations of the thermal evolution of sub-Neptune type planets,
up to ages of 5 Gyr \citep{lop13} show, however,   very little dependence
of radius on mass for a given H/He mass fraction and given incident stellar
flux. The radius decreases slowly with age after 1 Gyr, and the radius at
5 Gyr is practically independent of mass in the range 1--20 M$_\oplus$. The
radius at that time is very weakly dependent on stellar flux, 
but increases markedly with H/He mass fraction. These authors
therefore interpret the observed  trend of increased radius with mass as primarily
a composition effect, that is, on the average the higher mass planets have higher 
H/He mass fraction and therefore larger radii.

This paper investigates the origin and
evolution of such objects. The main issue is whether they formed {\it in situ}
or whether they formed at much larger distances, say 4--6 AU, and then,
or during the formation process, migrated inward to their present
orbital positions. Regarding the Kepler-11 planets, \citet{lis11}
question whether migration was involved because it would be expected
to result in      mean-motion resonances in multiplanet systems, which are
not observed in this particular system. However, later studies of 
dynamical and dissipative effects \citep{rei12,gol14}
show that in fact such planets can migrate though resonances, 
 thereby alleviating the constraint.
 Also, thermal evolution 
calculations for such planets \citep{lop12}, coupled with mass loss, indicate
that {\it in situ} formation is not likely and that the planets probably formed
beyond the snow line, with a substantial water component. 
 This conclusion
is based on models of \citet{iko12}         of the accretion of gas onto
already-formed rocky cores in the inner disk. The mass loss calculations
imply an inconsistency
for the Kepler-11 planets if they formed {\it in situ}: the original mass of
their H/He envelopes would have been larger than the amount they could have
accreted according to \citet{iko12}. 

Theoretical work by \citet{han12}
on the formation of sub-Neptunes and super-Earths assumes that they
form {\it in situ} but that much of the required solid material arrives by
migration of rock-sized objects from larger distance. This process was
originally investigated, in the context of hot Jupiters, by \citet{war97} and
applied to the case of 51 Peg b by \citet{bod00}. \citet{han12} find that gas
accretion onto rocky cores is likely to occur if the accumulated disk mass in solids
inside 1 AU is greater than about 25 M$_\oplus$, which is about 8  times that in 
the minimum-mass solar nebula (MMSN).
\citet{chi13} consider the formation of super-Earths {\it in situ} with no 
migration, based on a disk model that is somewhat enhanced in solid material
with respect to the minimum-mass solar nebula (see also \citet{han13}). 
Their rough estimates indicate that  rocky planets
in close orbits can capture and retain gaseous envelopes of a few percent
to tens of percent of the planet's mass. 

\citet{iko12} do {\it in situ} calculations, from the full stellar structure equations,
of the accretion of gas onto rocky cores in the inner disk
during the dissipation phase of the gas disk.
Parameters include the core mass $M_\mathrm{core}$, the disk temperature $T_\mathrm{neb}
$, and the characteristic disk dissipation time $\tau_d$.  Core masses range
from 1 to 10 M$_\oplus$. Calculations end when 
the disk has completely dissipated.  For example, for $M_\mathrm{core}=4$ M$_\oplus$, 
$T_\mathrm{neb}=550$ K, and  $\tau_d=10^5$ yr, the accumulated gas envelope mass
is about $2 \times 10^{-2}$  M$_\oplus$  if core cooling is not considered, and is
about $10^{-3}$  M$_\oplus$ in the more realistic case when  the cooling is included.
The range of ratios of envelope mass to
core mass is from 0.0002 to 0.1. Reasonably good agreement between these envelope
masses and those deduced for the Kepler-11 planets \citep{lop12,lis13} can be 
obtained for $\tau_d \approx 1$ Myr, but not for significantly shorter times.

In contrast, \citet{rog11}
calculate detailed core-accretion models for sub-Neptune objects forming 
 at 4 AU and 5.2 AU, assuming that they later migrate to positions where
the equilibrium temperature is 500 to 1000 K.  After accretion ends, the planets, 
which have  $M_\mathrm{core}$ in the range 2.5--4 M$_\oplus$, 
are evolved up to ages of 4 Gyr. The full evolutionary calculations are 
supplemented with static core/envelope models that  cover a wide range (1 to 20 M$_\oplus$)
 of planet masses 
$M_\mathrm{tot}$ and ratios $M_\mathrm{env}/M_\mathrm{tot}$. Theoretical radii  at late times from
these calculations agree,  generally for $M_\mathrm{env}/M_\mathrm{tot} < 0.1$, 
with those observed by {\it Kepler}.
The present work extends these formation calculations to a range of distances
from 0.5 AU to 4 AU and considers in further detail  the question regarding how much gas
can be accreted by  heavy-element cores
in the 2.2--2.5 M$_\oplus$ range, in the warm inner regions of the disk. The implications
regarding formation {\it in situ} or formation accompanied by migration are then discussed.

\section{Computational Method}

The calculations for these low-mass planets, with final core masses
of 2.2--2.5 M$_\oplus$, include two phases. 
The formation phase, starting with  $M_\mathrm{core} \approx 
$ 1 M$_\oplus$,  involves accretion of heavy-element ``core" mass as well
as gaseous envelope mass. The following evolutionary phase, starting at 
the end of accretion, involves 
 contraction and cooling of the gaseous envelope, at constant mass and with
the planet isolated from the disk, 
up to a final age of several Gyr. The  computational method and physical 
assumptions are
 described in detail in previous publications
\citep{pol96,mov10,rog11} and references therein.
During the early part   of the formation
phase, the gaseous envelope has low mass, and the core fairly
rapidly accretes to close to its final mass. 
The later parts of the formation phase are characterized by slow 
envelope accretion at practically constant $M_\mathrm{core}$. The core
accretion rate  is given by the standard equation \citep{saf69}
\begin{equation}
\frac{dM_{\rm core}}{dt} = \pi R^2_{\rm capt} \sigma \Omega F_g
\end{equation}
where $R_{\rm capt}$ is the effective geometrical
capture radius for planetesimals, $\sigma$ is the mass per unit area of
solid material (planetesimals) in the disk, $\Omega$
is the planet's orbital frequency, and $F_g$ is the
gravitational enhancement factor to the geometrical capture cross-section. 
The planetesimal
radius is taken to be 100 km, and $F_g$ is taken from \citet{gre92}.
In fact, if a reasonable distribution
of planetesimal sizes were taken into account, the formation time would
be reduced.
 However,  the large size partially compensates for the effect that 
the formation times based on  \citet{gre92} are somewhat faster
than those found in more detailed simulations, e. g., \citet{ina03}. In any case, 
the core grows rapidly in the inner region of a protoplanetary disk, and the
precise value of the planetesimal size has little effect on the outcome.
 The surface density of planetesimals changes with time according to the prescription
of \citet{pol96}, in which the accretion onto the protoplanet is taken 
into account, the feeding zone extends 4 Hill radii on either side of
the planetary orbit, and $\sigma$ is assumed to be uniform with radius
within the feeding zone. 

In general $R_{\rm capt} > R_{\rm core}$,
the radius of the heavy-element core, unless the envelope mass is 
negligible. 
In our  trajectory calculations, if the effects  on a planetesimal of 
gas drag,  ablation and fragmentation result in loss of over half of its
mass before it hits the core or escapes, 
it is considered to have been captured by
the envelope, and the amount of mass deposited in each layer is determined
\citep{pod88}. The effective  $R_\mathrm{capt}$  determined in this way 
(see \citet{pol96} for details)  can
be several times larger than $R_\mathrm{core}$.
The dusty material deposited in the envelope enters into the opacity
calculation (see below); it then is allowed to
sink to the core.  Thus the change in envelope composition caused by the 
deposition of solid material is not taken into account, although
\citet{hor11} show that an envelope enhanced in heavy elements can
significantly increase the gas accretion rate at a given core mass and can reduce the
critical core mass, that is, the mass required for rapid gas accretion to occur. 
However, \citet{iar07} show that 
 the organic and rock components of the planetesimals in fact do not
dissolve in the envelope and do sink, thus our assumption is valid for most
of our simulations, those inside the snow line (2 AU or less). 
Icy material however does dissolve,  so that in our calculations at 4 AU
 the gas accretion rate may be affected, and
 the ``core mass" is somewhat overestimated. What we call $M_\mathrm{core}$
there actually is the  mass   of  heavy elements in the planet, after
subtraction of the heavy-element  component of the nebular material accreted
at the surface of the envelope. 

The structure  and evolution of the gaseous  envelope is calculated 
according to
the standard spherically symmetric equations of stellar structure \citep{kip90},
augmented by the effects of the time-dependent $M_\mathrm{core}$, the
accretion rate of the envelope,  and the interactions of the incoming           
planetesimals with the envelope.  The temperature gradient is assumed to
be the adiabatic gradient in convection zones.
The main  energy sources are
planetesimal accretion, and contraction and cooling 
of the gaseous envelope.  In two cases, test runs were carried
out including heating effects resulting from radioactive decay in the core
 and from cooling of the 2.2 M$_\oplus$ core \citep{net11}. The radioactive
decay had practically
no effect on the radius at 4 Gyr.  The core heating is calculated, assuming an
isothermal core, according to
$ L_\mathrm{core} = -C_\mathrm{core} M_\mathrm{core} dT_c/dt$ where 
$ L_\mathrm{core}$ is the luminosity delivered to the envelope by the
cooling of the core, $C_\mathrm{core} \approx  1 \times 10^7$ erg g$^{-1}$ K$^{-1}$ is
the specific heat of the core, and $T_c$ is the temperature at the core/envelope
interface.  In agreement with \citet{lop13}, during the evolutionary (cooling) phase
this luminosity can contribute 25\% to 75\% of the total internal 
luminosity, depending on the time,  and it acts to lengthen the time to cool
to a given radius. However, with or without this additional
energy source, by the time the evolution reaches 4 Gyr, the radius is changing
very slowly, so the inclusion of $L_\mathrm{core}$ makes practically no difference on the 
final value. The test runs including  the core heating  resulted in increases
in the final radius of less than 5\%. On the other hand, during the accretion
phase, $T_c$ first increases with time as core and envelope increase in mass, so
the envelope actually delivers energy to the core. Later in the accretion phase
$T_c$ decreases, and this energy is returned to the envelope. In any case, this
luminosity is calculated to be only a few percent of the accretion luminosity and certainly
has little effect on the radius or envelope mass at the end of accretion. 
Therefore, the results quoted here for
final radii do not include these effects, as they fall below the overall
level of uncertainty in the simulations. The full set of equations 
is  solved by the Henyey method \citep{hen64}.

The radius at the inner boundary of the
envelope  is set to $R_\mathrm{core}$, which is determined
from its current mass. 
The core is composed either of iron and rock, with mass fractions 
30\% and 70\%, respectively,  or iron, rock, and ice, with mass fractions
10\%, 23\%, and 67\%, respectively, 
depending on the formation location. Given $M_\mathrm{core}$, the radius is
calculated from the equation of state of   \citet{sea07}, as summarized by
\citet{rog11}.
The equation of state  in the H/He envelope is taken
to be that given by \citet{sau95}, which includes            
the partial degeneracy of the electrons as well as non-ideal effects in the gas.
The chemical composition of the envelope is taken to be near-solar, with $X=0.70, 
~Y=0.283$, and $Z=0.017$, where $X,~Y,~Z$ are, respectively, the
mass fractions of H, He, and all remaining elements.

 The main feature of the Rosseland mean opacity during the formation phase 
involves the use of dust grain opacities that take into account
the settling and coagulation of the grains \citep{mov10}. 
Dust grains enter the envelope through the ablation of planetesimals
and are also carried in along with the accreting gas. The initial grain
size is 0.1 $\mu$m.  Grains are assumed to be spherical and to lack void
spaces.  The grain growth and settling are calculated
in detail as described in \citet{mov08} and \citet{mov10}.  Grain growth  up
to 2.58 mm in size is considered. The grain size distributions
and the Rosseland mean opacities are recalculated in every layer at every time step.
The grains are assumed to be composed of pure silicates, with a dust-to-gas ratio
of 1:100 by mass within the gas that is accreted by the planet.
At or  interior to 2 AU, this approximation is reasonable; even
at 4 AU the error introduced by this assumption is insignificant when
one considers the uncertainties in grain shape, sticking probability in collisions,
and radiative properties. The sticking probability is a parameter in the
grain code, and it is set to unity in the present simulations.
These opacities 
regulate the rate at which the envelope can contract, and therefore
affect the gas accretion rate.

Above 3000 K, the opacities of 
\citet{ale94} are used. Between 3000 K and 1800 K, the \citet{fre08} molecular 
opacities,
which do not include dust grains, are used. Below 1800 K, the dust opacities
of \citet{mov10} are added to the molecular opacities.
Once the planet has 
reached its final mass, as determined by the lifetime of the disk, 
the grains settle rapidly, evaporate in the interior, and are no longer
a significant opacity source. For
the final isolation  phase at constant mass, the molecular/atomic
opacities of \citet{ale94} and \citet{fre08} are used, with solar composition. 

The gas accretion rate
 is determined by the condition  that the planet outer  radius $R_p \approx 
R_\mathrm{eff}$, where 
the effective accretion radius is given by
\citep{lis09}           
\begin{equation}
R_{\rm eff} = \frac{GM_\mathrm{tot}}{c_s^2 + \frac{GM_\mathrm{tot}}{KR_H}}~.
\label{eq:reff}
\end{equation}
Here $c_s$ is the sound speed in the disk, $R_H$ is the 
Hill sphere radius, and $M_\mathrm{tot}$ is the total mass of the planet.
The constant $K \approx 0.25$ is determined by three-dimensional
numerical simulations of disk flow in the vicinity of an embedded 
planet  \citep{lis09}. These simulations show that the planet does
not retain gas from the entire Hill-sphere volume; rather, it can
occupy a region with radius $\approx 0.25 R_H$.
Thus, if $R_H$ is  small compared with the Bondi
accretion radius $R_B=GM_\mathrm{tot}/c_s^2$, $R_{\rm eff}= 0.25 R_H$.

The surface boundary conditions depend on whether the planet is
still accreting (the formation phase) or whether it is isolated (the
evolutionary phase). During the formation phase the
temperature is  set to a constant 
value appropriate for the protoplanetary disk, 
$T_\mathrm{neb}$ (see Table \ref{table:1}). These 
temperatures correspond approximately to a ratio
of sound speed to orbital speed of 0.05, except at
0.5 AU, where a slightly smaller ratio is taken.  The density
at the surface, $\rho_\mathrm{neb}$, 
is determined initially from $\rho_\mathrm{neb} = \sigma_g/2H$, where 
$\sigma_g$ is the gas surface density, the scale height  $H=0.05 a_p$, and
$\sigma_g/\sigma =70$ at 4 AU and 200
at 2, 1, and 0.5 AU
($a_p$ is the distance of the planet from the star). In three of the runs, the
disk lifetime is arbitrarily set to 2 Myr.  The density $\rho_\mathrm{neb}$
 is assumed to decline
linearly with time up to 1.9 Myr; then it is cut off more rapidly, to near zero,
on the time scale of 10$^5$ years. Three other runs have the disk cutoff time
set to match a pre-chosen envelope mass.  Here also the density drops
rapidly during the 10$^5$ years just before cutoff. 
Envelope masses remain low in all calculated
cases, so that the phase of rapid gas accretion associated with the
growth of Jupiter-mass planets  never occurs.

The isolation
mass for the heavy-element planetary core, which in these simulations
turns out to be close to the final core mass, is given by 
\begin{equation}
M_\mathrm{iso} = \frac{8}{\sqrt{3}}(\pi C)^{3/2} M_\star^{-1/2} 
\sigma^{3/2} a_{p}^3~,
\label{eq:iso}
\end{equation}
where  
$M_\star$ is the mass of the central star, and  
$C \approx 4$, the number of  Hill-sphere radii defining the region, 
on each side of the planetary orbit, 
from which the object  is able to capture planetesimals \citep{lis87}. 
Once $M_\mathrm{core} \approx  M_\mathrm{iso}$,
the $dM_\mathrm{core}/dt$ slows down drastically, but  gas accretion continues.
In these low-mass models the crossover mass
 (when $M_\mathrm{core} = M_\mathrm{env}$)
is never reached. Note that the calculation method implicitly assumes
that  the core mass will reach $M_\mathrm{iso}$; formation of multiple
embryos in the vicinity of $a_p$ is not considered.

When  the disk becomes very tenuous near the end of its lifetime,
the planet makes a transition to isolated boundary conditions
that  take into account the radiation effect of the central star.
Gas and solid accretion stop, and Equation (\ref{eq:reff}) no longer applies. 
Then 
\begin{equation}
L_\mathrm{tot} =4 \pi R_p^2 \sigma_B T_{\rm eff}^4 ~~~~~{\rm and}~~~~~~
\kappa_\mathrm{R} P = \frac{2}{3} g~,
\end{equation}
where $\sigma_B$ is the Stefan-Boltzmann constant, $T_{\rm eff}$
is the surface temperature, $L_\mathrm{tot}$ is the total luminosity which
includes the internal luminosity from the planet and the re-radiated
stellar input, 
and $\kappa_\mathrm{R}$, $P$, and $g$
are, respectively, the photospheric values of Rosseland mean opacity, 
 pressure, and acceleration of gravity.
The surface temperature is calculated from 
\begin{equation}
T^4_\mathrm{eff} = T^4_\mathrm{int} + T^4_\mathrm{eq}~,
\end{equation}
where $T_\mathrm{int}$ is determined    by the internal luminosity and
the outer radius of
the planet (a generally small contribution), and $T_\mathrm{eq}$, 
the equilibrium temperature of the planet in the radiation field of
the star, is taken to be a parameter.

\section{Calculations and Results} \label{sec:results}

The main parameters that are varied from run to run  are the  formation position of the
planet in the disk, the value of the initial solid surface density 
$\sigma$ at that position, 
 and the time for cutoff of gas accretion.  The planet's
core mass is determined through the calculation itself;  in fact it is
practically fixed by the choice of $a_p$  and $\sigma$, which determine $M_\mathrm{iso}$ 
(Equation \ref{eq:iso}). $M_\mathrm{iso}$ is taken to be just above  2 M$_\oplus$, in
order to provide a comparison with Kepler-11 f and other planets with
core mass in that range.
The formation process is assumed to
take place at a fixed orbital radius.  The initial value of the core mass
is about 1 M$_\oplus$, and the starting time is arbitrarily set 
to $2 \times 10^5$ yr, representing the approximate time to build this
core  at 5.2 AU with $\sigma=10$ g cm$^{-2}$ \citep{lis09}. 
Closer to the star  this time will be shorter, but in these calculations
the total time to build the core is short compared with the overall
evolutionary time. After formation, the planet
is assumed to take on three different insolation temperatures: 
250 K, 500 K, and 1000 K. In most cases this assumption implies a
modest amount of migration. (Inward migration, were it to occur during gas accretion,
would have the effect of reducing the final $M_\mathrm{env}$, because
of the gradual reduction in $R_B$ in the disk.) The radii at the end of a 4 Gyr constant-mass
evolution are determined for each of these temperatures.

\begin{table}
 \caption{Input Parameters and Results}\label{table:1}
 \centering
  \vskip 0.2 in%
 \resizebox{1.0\textwidth}{!}{%
 \small
 \begin{tabular}{|l||ccccccc|}
 \hline\hline
 Run $\rightarrow$
 &  4H  & 4 & 2H  & 2
 & 1  &  0.5
   \\
 \hline
Distance from star (AU) 
 &   4   & 4  & 2  & 2  & 1  & 0.5
  \\
 \hline
Disk temperature (K) 
 &   126    & 126   & 310    &   310   &  650    &  927
    \\
 \hline
H$_2$O in solid core & Y & Y & N & N & N & N
   \\
 \hline
Disk solid $\sigma$ (g cm$^{-2}$) 
 &    6  &   6  &   22   &   22   &   90  &   360
   \\
 \hline
MMSN solid $\sigma$ (g cm$^{-2}$) 
 &    3.8  &   3.8  &   2.5   &   2.5   &   7.0  &   20
   \\
 \hline
Disk gas  $\sigma$ (g cm$^{-2}$)
 &  420  &  420  &  4500  &   4500 &  18000   &   72000
    \\
 \hline
Accretion cutoff (Myr)           
 &    0.64   &  0.49   &  2.0   &   0.42     &   2.0   &    2.0
   \\
 \hline
Final $M_\mathrm{core}$  (M$_\oplus$)
 &    2.50  & 2.44  &  2.20  &   2.15  &   2.20    &  2.20
    \\
 \hline
Final $R_\mathrm{core}$  (R$_\oplus$)
 &    1.65  & 1.64  &  1.22  &   1.21  &   1.22    &  1.22
    \\
 \hline
Final $M_\mathrm{env}$  (M$_\oplus$)
 &   0.16   & 0.054 & 0.16   &   0.054  &   0.054    &  0.037 
    \\
 \hline
$M_\mathrm{env}/M_\mathrm{tot}$
 &    0.060  &  0.022  &  0.068   &   0.024    &   0.024   &    0.017
   \\
 \hline
Radius (4,250)  (R$_\oplus$)                  
 &   2.72   &  2.19   &  2.41   &   1.81     &   1.82   &    1.69
   \\
 \hline
Radius (4,500)  (R$_\oplus$)                  
 &   3.24   &  2.64   &  2.88   &   2.19     &   2.19   &    1.89
   \\
 \hline
Radius (4,1000)  (R$_\oplus$)                  
 &   6.18   &  4.51   &  5.79   &   3.91     &   3.94   &    3.13
   \\
 \hline
 \end{tabular}
       }
\end{table}

The parameters  and basic results for the runs are given in Table 
\ref{table:1}. The column headings 
in the table give the run identifiers; the six runs are labelled 
4H, 4, 2H, 2, 1, and 0.5. The numeral gives the distance from a 1 M$_\odot$
star in AU. The symbols 4H and 2H indicate that the final envelope
mass has the higher value of 0.16  M$_\oplus$,   while  in Runs 4 and 2 accretion was 
terminated when the final $M_\mathrm{env}$ reached the lower value 0.054  M$_\oplus$.
The first seven rows below the run identifiers  in the table give assumed parameters: $a_p$, 
$T_\mathrm{neb}$, core composition (Y indicates the presence of ice; 
N indicates no ice),  $\sigma$, the corresponding $\sigma$ in the MMSN,
 the initial gas surface density $\sigma_g$, and the assumed gas accretion
cutoff time. \emph{Note that our assumed final core masses in the range 
2 M$_\oplus$ require very high values of $\sigma$ in the inner disk}, as 
determined by Equation (\ref{eq:iso}). The bottom seven rows  give results: the final values of
$M_\mathrm{core}$, $R_\mathrm{core}$, $M_\mathrm{env}$, the final ratio of
 $M_\mathrm{env}$ to the total mass, and three final values of the
radius $R (4,n)$, where the 4 refers to the final time of about 4 Gyr, and
$n$ gives the assumed value of $T_\mathrm{eq}$.  These radii
span    the 1.7--6 R$_\oplus$ range  that  encompasses most of the planets
observed by {\it Kepler}.

The results presented in Table \ref{table:1} show that while the
envelopes accreted by 2.2--2.5 M$_\oplus$ cores are of low mass, they
occupy the majority of the planetary volume  even after cooling for 4 Gyr at
distances from their star where $T_\mathrm{eq}$ is only 250 K. For a given run, 
 that is, for a given mass and time,
\emph{planetary size increases modestly with $T_\mathrm{eq}$},  as also found by
\citet{lop13}.
For planets with the same $M_\mathrm{core}$ and the same accretion
cutoff time (Runs 0.5, 1, and 2H), the final $M_\mathrm{env}$ increases
with distance from the star, mainly as a result of the decreasing disk 
temperature  and increasing $R_H$ and $R_B$ with distance.   This result is
at least qualitatively consistent with that of \citet{iko12}. Also,   as previously
emphasized by \citet{lop13}, there is
a noticeable increase in radius, given $T_\mathrm{eq}$,  for increasing 
$M_\mathrm{env}$.   Other differences are more subtle.

Runs 4H and 2H have the same value of $M_\mathrm{env}$, but
at a given  $T_\mathrm{eq}$ the 4H planets are larger. This result occurs
 because the core of Run 4H, which includes ice, 
is of lower density and larger size. 
Thus, despite the core being slightly more massive in Run 4H, 
 the pressure at the base of the envelope is lower, leading to a
lower density. For the $T_\mathrm{eq}=250$ K case, the pressures are  
$3.35 \times 10^{10}$ and $6.91 \times 10^{10}$ dyne cm$^{-2}$, 
 respectively, for 4H and 2H, and the corresponding densities are
0.17 and 0.23 g cm$^{-3}$. The differences between Runs 4 and  2 are 
explained similarly.
The differences in the final radii between Runs 1 and 2, which have the
same $M_\mathrm{core}$, $R_\mathrm{core}$,  and $M_\mathrm{env}$, are small. 
Despite differences in initial conditions, one would expect these two cases
to converge to the same radii after 4 Gyr, because Kelvin-Helmholtz times
for the envelopes are  relatively short ($\approx 10^8$ yr). 
The small differences ($<1$\%) are well within the
uncertainties of the simulation, which are $\approx 20$\% in the radius at  
$T_\mathrm{eq}=1000$ K, $\approx 10$\% at $T_\mathrm{eq}=500$ K, and a few percent at 
$T_\mathrm{eq}=250$ K \citep{rog11}.
The uncertainty arises because molecular opacities \citep{fre08} have not been tabulated  
for some of the pressures encountered in the interior; they have to be
extrapolated. The opacities  determine the location of the boundary between the outer
radiative zone and the inner convective zone, which has an effect on the radius.
The uncertainties are systematic; thus the differences are  smaller
than the uncertainties.

\begin{figure}[ht]
\centering%
\includegraphics[angle=0,height=90mm]{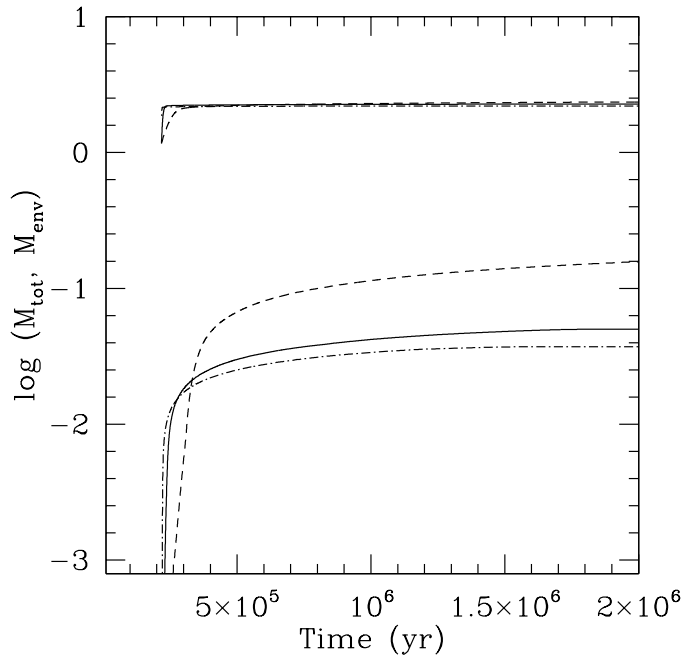}
\caption{Upper curves: total mass (in Earth masses); lower curves: envelope
mass (in Earth masses), both  as a function of time during the formation
phase for Runs 2H, 
1, and 0.5, all of which have the same core mass and the same gas cutoff
time of about 2 Myr. 
{\it Solid curves:} Run 1; {\it dashed curves:} Run 2H; 
 {\it dash dot curves:} Run 0.5.             
             }
\label{fig:1}
\end{figure}
\begin{figure}
\begin{center}
\includegraphics[angle=0,height=90mm]{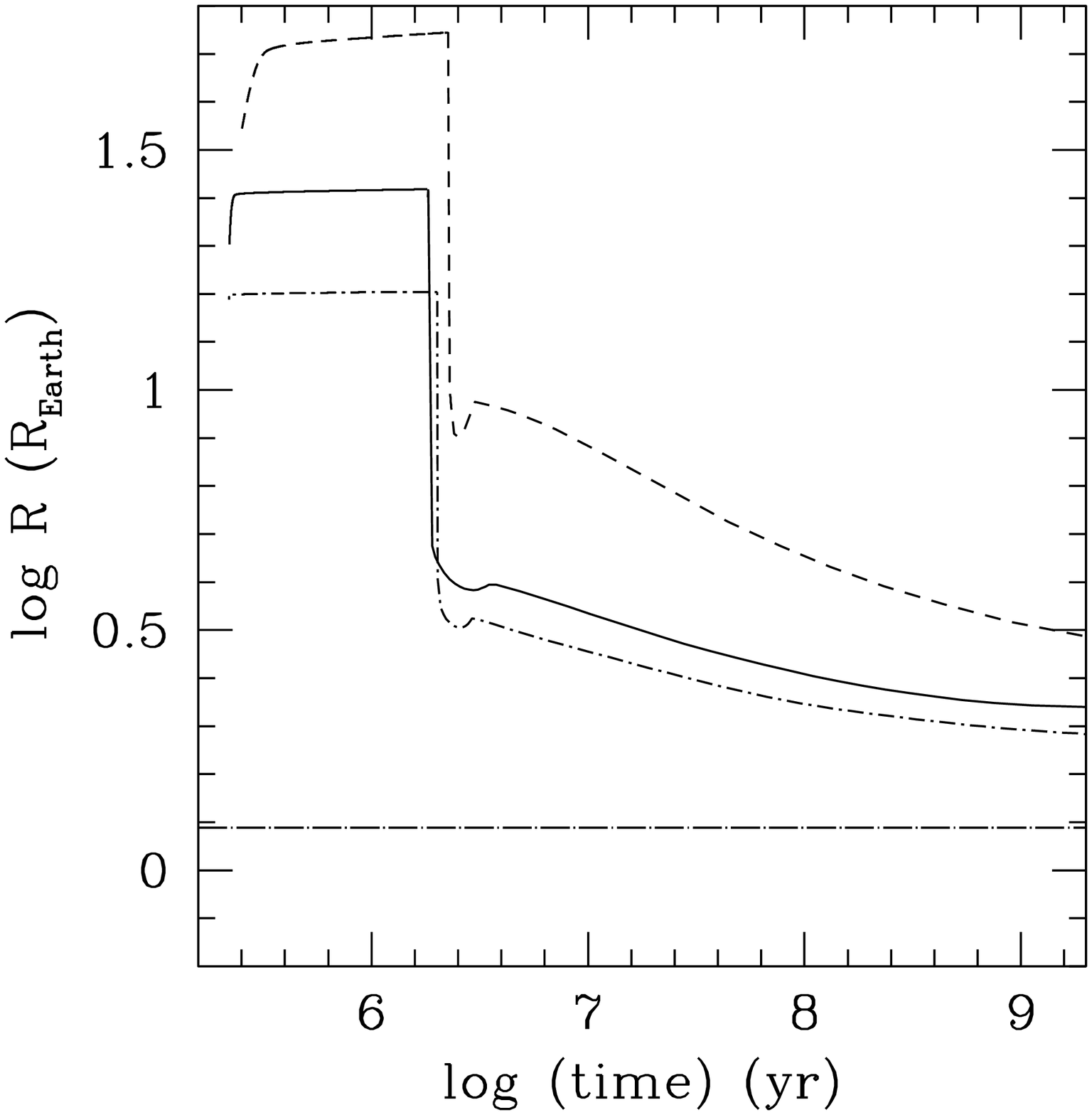}
\caption{
Evolution of the radii of sub-Neptune type planets forming at
2 AU ({\it dashed curve}), 1 AU  ({\it solid curve}), and 0.5 AU
 ({\it dash-dot curve}); Runs 2H, 1, and 0.5, respectively.
 The final mass of the rock-iron core
is 2.2 M$_\oplus$ in all cases, and  the corresponding radius, as 
obtained from the equation of state of \citet{sea07}, is shown as the
{\it long-dash dot curve}. The sharp decrease in radii occurs during
the transition from the formation phase to the constant-mass evolution
phase. During the evolution phase, the equilibrium temperature is
$T_\mathrm{eq}=500$ K.  The slight increase in the radii at about 
3 Myr is a result of the transition to that temperature. 
}
\label{fig:2}
\end{center}
\end{figure}

The buildup of the mass during the formation phase for Runs
2H, 1, and 0.5 is shown in Figure \ref{fig:1}. Note that the
 core mass is very close to its final value (2.2 M$_\oplus$)  less than
 $10^5$ yr after the starting time, with the time required to reach 90\%
of the final core mass  decreasing
with decreasing distance from the star. Envelope masses increase on a time scale
of 1 Myr; after that time the rate of increase is very slow.
At the later times the addition of mass to the envelope is
driven by a very slow contraction, since the value of $R_\mathrm{eff}$
is practically constant. Just before 2 Myr the values of 
$\dot M_\mathrm{env}$ are $5 \times 10^{-8}$, $3 \times 10^{-9}$, 
and $4 \times 10^{-10}$ M$_\oplus$ yr$^{-1}$, respectively, for
Runs 2H, 1, and 0.5.  If the runs were to be continued for an
additional  Myr, the increases in $M_\mathrm{core}$ in all cases
and in $M_\mathrm{env}$ for Runs 1 and 0.5 would be 
negligible. In Run 2H the envelope would be expected to accrete
an additional $\leq  0.05$ M$_\oplus$.

The final $M_\mathrm{env}$ increases with
increasing distance from the star, a result of the increase
in $R_H$ and $R_B$. The factors that determine these masses
are complex. It	turns out that for these three cases, in each
case $R_B$ and 0.25 $R_H$ agree to within about 20\%. The
values of $R_\mathrm{eff} \approx 0.5 R_B$ are $1.0 \times 10^{10}$, 
$1.7  \times 10^{10}$, and $3.5 \times 10^{10}$ cm, for Runs
0.5, 1, and 2H, respectively. The envelope masses are found to
scale roughly as $R_\mathrm{eff}^3 \rho_\mathrm{phot}$, where
$\rho_\mathrm{phot}$ is the photospheric density (where the inward-integrated
optical depth approaches unity)  as determined
from the detailed models at the time when the envelope has obtained
most of its mass. Note that $\rho_\mathrm{phot}$, which depends on the
details of the run of opacities in the model, and which is a modestly
decreasing function of distance from the star, is the appropriate
scaling factor, not $\rho_\mathrm{neb}$. The outer regions of the 
model, which have negligible mass and which are optically thin, are
not important.  For Runs 2H, 1, and 0.5 the final envelope masses
are respectively 7\%, 2.4\%, and 1.7\% of the total mass. The actual
values of $M_\mathrm{env}$ scale as 1:~0.34:~0.23 while the corresponding
values of $R_\mathrm{eff}^3 \rho_\mathrm{phot}$ scale as 1:~0.32:~0.27.

The radii as a function of time for the same three cases are shown in
Figure \ref{fig:2}. During the accretion phase, these radii fall just
below $R_\mathrm{eff}$, which is approximately constant once $M_\mathrm{core}$
has approached its final value of 2.2 M$_\oplus$. At the age of 2 Myr, the disk
dissipates and the radii fall rapidly as a result of the transition to
isolated (irradiated photospheric) boundary conditions. During the evolution
phase up to 4 Gyr, the radii decrease gradually to final values in the
range 2--3 R$_\oplus$ for  $T_\mathrm{eq} = 500$ K. Run
2H ends up with a larger radius than in the other two cases because of its
higher envelope mass. 

\begin{figure}[ht]
\centering%
\includegraphics[angle=0,height=80mm]{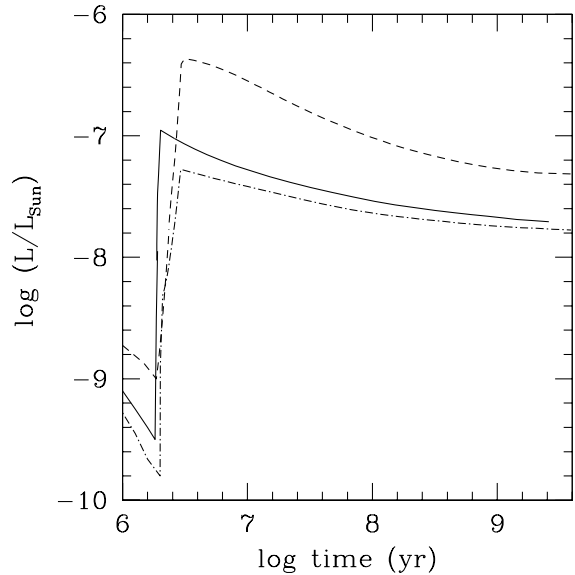}
\caption{ Total luminosity (in solar units), including intrinsic as
well as insolation effects,  as a function of time for Runs 2H, 
1, and 0.5, all of which have the same core mass and the same gas cutoff
time, 2 Myr. The plot starts near the end of the accretion
phase, during which insolation effects are not included. Once the planet
becomes isolated, the assumed value of $T_\mathrm{eq}$ in all cases
rises to  500 K  and the luminosity rises sharply. The slow decline in
the later phases results from contraction at constant  $T_\mathrm{eq}$. 
{\it Solid curve:} Run 1; {\it dashed curve:} Run 2H; {\it dash dot curve:} Run 0.5.}             
\label{fig:3}
\end{figure}

\begin{figure}[ht]
\centering%
\includegraphics[angle=0,height=90mm]{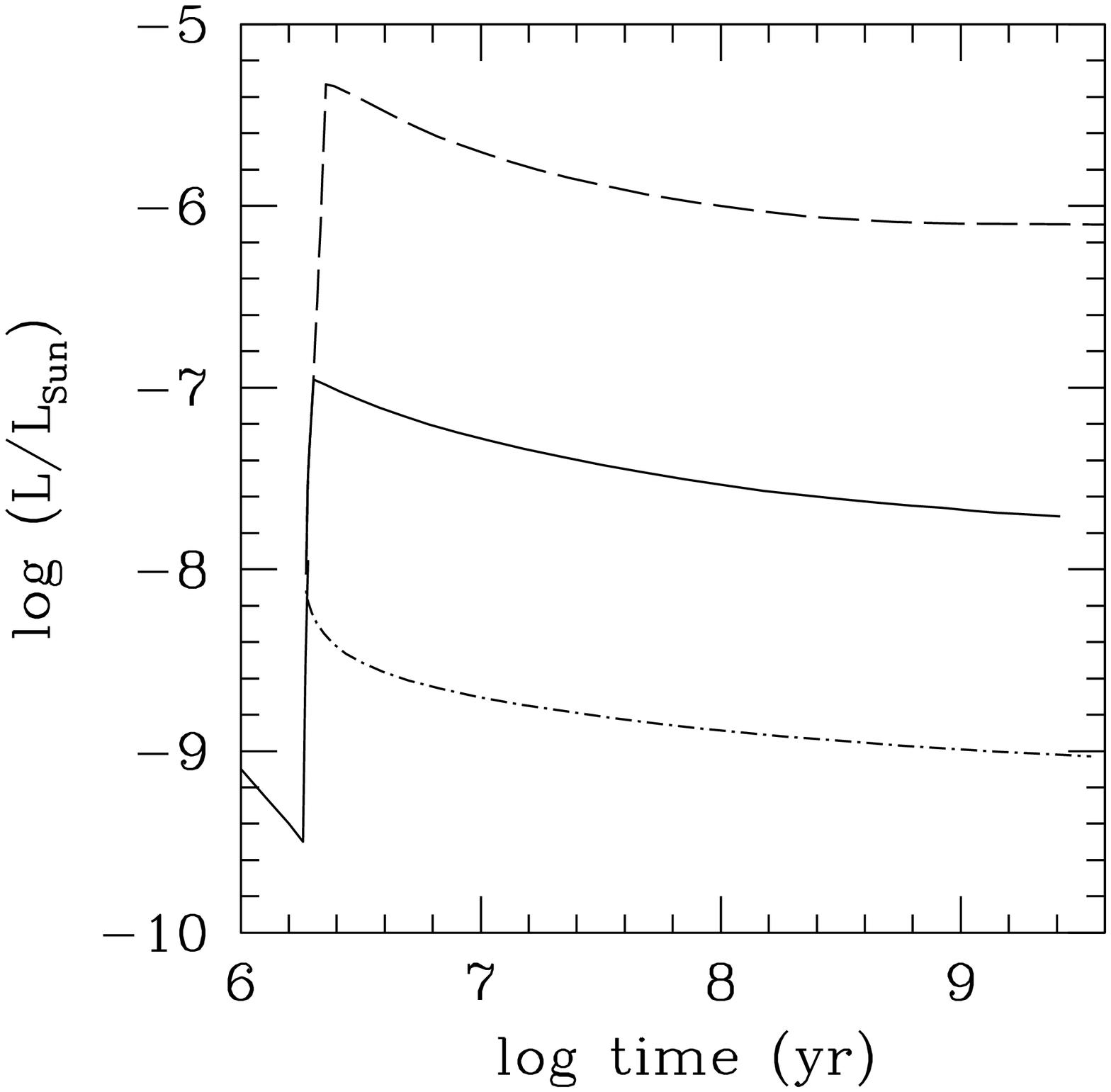}
\caption{Total luminosity (in solar units), as in Figure \ref{fig:3},
 as a function of time for Run 1, 
illustrating the effect of the assumed value of $T_\mathrm{eq}$. 
{\it Solid curve:} $T_\mathrm{eq}=500$ K; 
 {\it dash dot curve:} $T_\mathrm{eq}=250$ K;
 {\it dashed   curve:} $T_\mathrm{eq}=1000$ K.
             }
\label{fig:4}
\end{figure}
The end of the formation  phase  and the entire 
constant-mass evolutionary phase of
Runs 2H, 1, and 0.5 are illustrated in Figure
\ref{fig:3}, which gives the total luminosity as a function of time. 
The luminosity at the end of the formation phase is the
intrinsic radiation from the planetary interior. After the time
of 2 Myr the accretion ends and the objects make a transition to
photospheric boundary conditions with insolation at $T_\mathrm{eq}=500$
K.  The luminosity is then dominated by insolation effects.
The higher luminosity in Run 2H as compared with the other cases is
a result of its larger radius.  Figure \ref{fig:4} shows a similar plot
for the case of Run 1, with three different values of $T_\mathrm{eq}$
during the evolution phase. The separation of over an order of magnitude
in luminosity between the curves represents the effect of 
 $T_\mathrm{eq}^4$ as well as the increased radius as a function of  $T_\mathrm{eq}$.

\begin{figure}[ht]
\centering%
\includegraphics[angle=0,height=90mm]{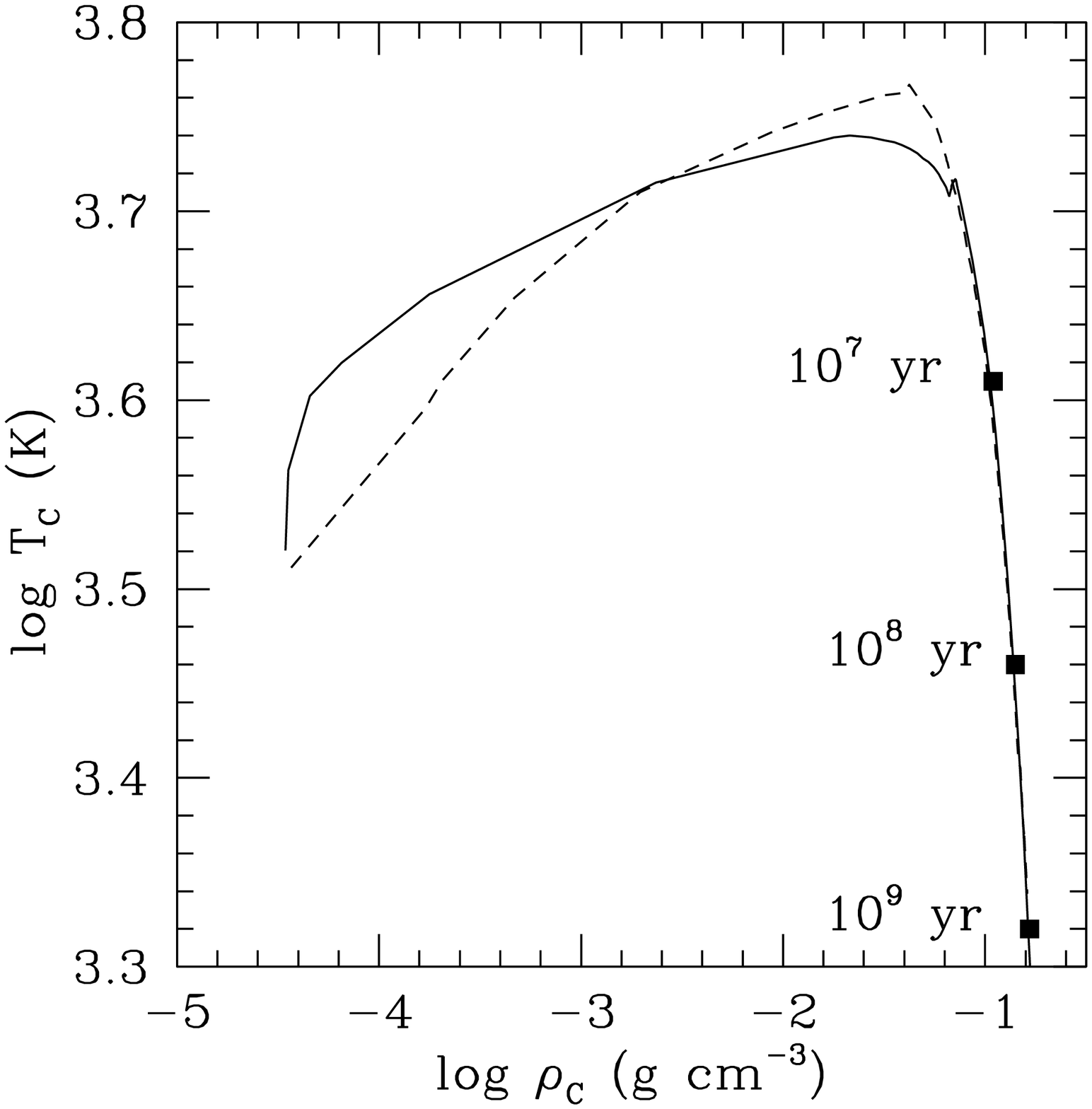}
\caption{Evolution with time of the 
central temperature (at the core/envelope interface) as a 
function of central density for Run 1 {\it (solid curve)} and Run 2 
{\it (dashed curve)}. During the isolation phase both planets have $T_\mathrm{eq} = 500$ K.
 Elapsed times during the final cooling phase
are indicated by labelled solid squares. 
             }
\label{fig:5}
\end{figure}

Figure \ref{fig:5} compares Runs 1 and 2, which have the same final
core masses and envelope masses, but were formed under different initial
conditions. The accretion for Run 2 is cut off (bifurcated from Run 2H)
at 0.42 Myr, because at that 
time its envelope mass matches that of Run 1. The figure shows
the evolution of density $\rho_c$ and temperature $T_c$, both evaluated at the
base of the envelope. Initially, Run 1, at 1 AU, accretes core mass faster
than does Run 2, at 2 AU. Therefore $T_c$ increases more
rapidly in Run 1. Run 1 reaches a maximum $T_c = 5500$ K at 
$2.82 \times 10^5$ yr, 
at which time the envelope mass is 0.016 M$_\oplus$. The slight 
secondary maximum in $T_c$, where log $\rho_c=-1.1$, occurs when gas accretion
cuts off at about 2 Myr. Beyond that point, the interior cools continuously.
In Run 2, once the core mass has levelled off close to its final value, the
central temperature catches up to that in Run 1 and reaches a
slightly higher maximum of  $T_c=5850$ K, at a time of 
 $4.2 \times 10^5$ yr, just before 
 the time of envelope accretion cutoff. The main reason for the 
more rapid temperature increase is that the envelope
accretion rate in Run 2 is considerably faster than that in Run 1 because of
the larger Bondi radius, leading to faster compression.
After cutoff, the evolutionary track joins that for Run 1, and
the two curves are practically identical after $10^7$ yr. These two curves
were calculated with $T_\mathrm{eq}=500$ K,  and both reached 
final radii of about 2.2 R$_\oplus$.
 
Runs 4H  and 4 incorporate the effect of forming the planet beyond the 
snow line at 4 AU at 	a disk temperature of 126 K. The isolation mass is
2.42 M$_\oplus$ under these conditions, with a value of $\sigma=6$ g cm$^{-2}$, 
less than twice that in the MMSN.            The core accretes to close to
$M_\mathrm{iso}$  at an age just under 0.5 Myr. In Run 4H the gas accretion is cut off
arbitrarily at 0.64 Myr with an $M_\mathrm{env}=0.16$ M$_\oplus$ and a total
mass of 2.66 M$_\oplus$. Had gas accretion been allowed to continue up to
2 Myr, as in Run 2H, the $M_\mathrm{env}$ would have been 0.54
 M$_\oplus$, as found by \citet{rog11}, not sufficient to
reach rapid gas accretion. This value of $M_\mathrm{env}$ is found
again to scale as $R_\mathrm{eff}^3 \rho_\mathrm{phot}$, where 
$R_\mathrm{eff} \approx 10^{11}$ cm. In the case of Run 4, 
the gas accretion is cut
off at 0.49 Myr, with  $M_\mathrm{env}=0.054$ M$_\oplus$ and a total mass of
2.494  M$_\oplus$. This envelope mass matches those in Runs 1 and 2.
Run 4  has a slightly larger core mass than does Run 2, and this pair of runs have
similar cutoff times    and exactly
the same envelope mass. The main difference between the two cases is that
the core in Run 4 is 30\% larger in radius, because it contains an ice
component.

\begin{figure}[ht]
\centering%
\includegraphics[angle=0,height=90mm]{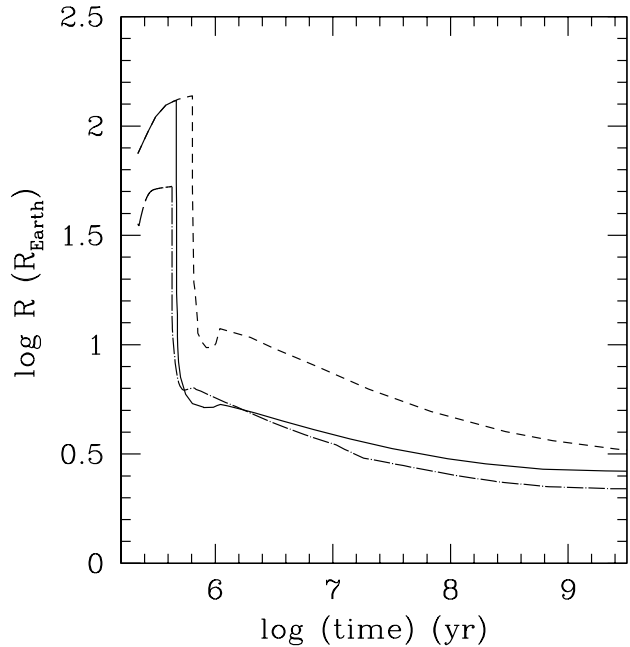}
\caption{Evolution with time of the 
radii (in units of R$_\oplus$) for Runs  4H {\it (dashed line)}, 
4 {\it (solid line)}, and 2 {\it (dot-dashed line)}. During the evolution
phase at constant mass, the value of $T_\mathrm{eq}$ is set to
500 K.
             }
\label{fig:6}
\end{figure}

Figure \ref{fig:6} shows radii as a function of time for Runs 4H, 4, and 2, 
all with $T_\mathrm{eq}=500$ K. The initial increase in $R_\mathrm{eff}$ in
Runs 4H and 4 corresponds to the growth of the core from  1 to 2.5 M$_\oplus$.
At cutoff time the radii decrease rapidly during the transition to
isolated boundary conditions. A small secondary maximum occurs just after
1 Myr as $T_\mathrm{eq}$ increases to 500 K. The  difference in the
final radii at 4 Gyr occurs because of the difference of a factor of 3 in
$M_\mathrm{env}$. Run 2 has a smaller maximum value of $R_\mathrm{eff}$
than in the other cases because of the smaller values of $R_B$ and $R_H$ at
its smaller distance. The noticeable difference in final radii between
Runs 2 and 4 is a result of the differences in core radii,  which influence the
pressure and thus the density in the lower envelope, even though
the core masses are about the same.

\section{Comparison with Kepler-11 f}

The transiting planet Kepler-11 f has a radius of 2.48 (+0.02,-0.03) R$_\oplus$,
$T_\mathrm{eq} \approx 525$ K, and the planet orbits at 0.25 AU
from its solar-type star. The  mass measured through
transit timing variations
is 2.0 (+0.8,-0.9) M$_\oplus$ \citep{lis13}, which allows it to be compared with
our theoretical models.   The age of Kepler-11 is estimated to be
8.5 (+1.1,--1.4) Gyr \citep{lis13}. Although the final radii from
our models (Table 1) are given at 4 Gyr, the evolution
to longer times shows a negligible decrease in radius between
4 and 8 Gyr, certainly within the overall uncertainty in the
simulations.

Our results  indicate that if the planet
had formed at 1 AU its present radius would have been 2.19 R$_\oplus$; if 
it had formed at 0.5 AU, the radius would have been 1.89 R$_\oplus$. 
Thus, if it had formed {\it in situ} at 0.25 AU, the radius would
have been even smaller. At that distance, it would not have
been able to accrete enough gas to account for its present
radius.  Thus formation {\it in situ} is not likely unless, 
possibly, its mass came out to be close to the one-standard-deviation 
upper limit from
the observations.  It is true that the actual disk lifetime could be 
longer than the 2 Myr we have assumed. However, the envelope accretion
rate at the end of the accretion phase for Run 0.5 was only
$4 \times 10^{-10}$ M$_\oplus$ yr$^{-1}$; thus even after an additional
5 Myr of disk lifetime the added envelope mass would have been
at most 0.002  M$_\oplus$, implying a negligible increase in radius.
The effect would be even smaller at 0.25 AU.
Furthermore, formation at 0.25 AU would imply a
nebular temperature of over 1000 K, where the value of $R_B$ is smaller
than at 0.5 AU,   and the required $\sigma$ would have to be $\approx 
1440$ g cm$^{-2}$, 25 times that in the MMSN, assuming that $M_\mathrm{core}
\approx M_\mathrm{iso}$.  
We consider it more  likely that the planet formed at a larger distance and
then (or during formation) migrated inward.  The actual formation location
could be either inside or outside the ice line, depending on the specific
parameters of the situation and the amount of mass loss sustained by the
envelope after formation.

Arguments in favor of  {\it in situ}
formation of super-Earth type planets inside 0.5 AU are given by
\citet{chi13}. Their estimate of the required $\sigma$ for
the Kepler-11 system is of the same order as ours.  Their estimate of
the envelope mass that can be accreted by Kepler-11 f {\it in situ}, 
for $M_\mathrm{core} = 2.2 
$ M$_\oplus$, is $M_\mathrm{env}/M_\mathrm{core} \sim  0.024$, slightly
higher than the value indicated by our more detailed simulations but
still not quite sufficient to account for the observed radius. At
 $M_\mathrm{core} = 2.8$ M$_\oplus$, however, their estimate would
be consistent with the observations.

\citet{iko12} considered gas accretion onto  the Kepler-11 planets
{\it in situ}, although the rocky cores were assumed to have migrated
inwards from their actual formation locations. 
For  parameters similar
to ours---namely $M_\mathrm{core} = 2.2$
M$_\oplus$, $T_\mathrm{neb}=550$ K, and $\rho_\mathrm{neb}$
enhanced by a factor of 10 relative to the minimum-mass
disk---they find $M_\mathrm{env}/M_\mathrm{tot} \approx 0.001$,
assuming a characteristic disk dissipation time  $\tau_d$ of $10^5$ yr.
If  $\tau_d$ is increased to 1 Myr, the value of 
$M_\mathrm{env}/M_\mathrm{tot}$ goes up to about 0.003. Their results
for the mass ratio 
are about 1 order of magnitude below ours. As a result of the low
envelope masses, their models do not provide a satisfactory fit to 
Kepler-11 f. There are several differences in the assumptions in their
calculations in comparison with ours.  First, they assume that
disk dissipation and envelope accretion are occurring simultaneously,
while in our case most of the envelope accretion occurs prior to
significant disk dissipation. Second, they include heat loss from
the rocky core in the envelope calculation, which we do not consider.
 As mentioned above, their final envelope mass depends on whether
or not core heating is included, and it depends on
the details of disk dissipation, while in our case, because envelope
accretion occurs much earlier, the core luminosity has little effect.
Third, they do not include planetesimal accretion as an energy
source in the envelope, an effect which we do include. Fourth,
differences in the details of the assumed  disk dissipation  procedure
could account for some of the differences in the results for the
envelope mass. Fifth, except in one case they do not include dust grains in the
calculation of the envelope opacity during the accretion phase.

Mass loss may have played a modest role in the post-formation
evolution of Kepler-11 f. Estimates in the literature of mass loss
rates are based on energy-limited hydrodynamic escape driven by stellar
EUV and X-ray radiation \citep{mur09}. Deduced histories of stellar
EUV and X-ray fluxes indicate that most of the mass loss occurs
at ages $<0.1$ Gyr \citep{rib05}. \citet{iko12} calculate a loss of
0.1 M$_\oplus$ for Kepler-11 f at its present orbit over its lifetime.
However,  they assume a value $\epsilon=0.4$, where $\epsilon$ is the fraction
of incident radiation that goes into driving mass loss. A more
usual value of 0.1 \citep{lop12} would result in a loss of about
0.025 M$_\oplus$ from the envelope. In either  case, the planet should
have lost its entire envelope according to their {\it in situ} calculation.
In the case of our estimated envelope mass at 0.25 AU, most or all
of the envelope would have been lost.

\citet{lop12} couple the mass loss
calculation with post-formation thermal evolution for the Kepler-11 planets
at their current stellar input flux levels.
 For Kepler-11 f, with
an assumed rock/iron core and gaseous H/He envelope,  and with the
preliminary mass estimate of 2.3(+2.2, --1.2) M$_\oplus$, they conclude that
the planet originally had roughly 30\% of its mass in the H/He
envelope, implying a mass loss of about 1 M$_\oplus$. A roughly similar
result was obtained under the assumption that the planet's
core contained water along with the rock and iron. This amount of mass
is clearly inconsistent with the envelope mass the planet could
have accreted by formation at its current orbit. 
According to our models,
in fact, a high  envelope mass, for $M_\mathrm{core}=2.2$ M$_\oplus$,
 is inconsistent with formation inside
the snow line, out to 2 AU (Run 2H). 

To estimate whether  the formation
of a planet with $M_\mathrm{core} \approx 2.5$ M$_\oplus$ and $M_\mathrm
{env} \approx 1.0$ M$_\oplus$ is possible outside the snow line,
we extend the Run IIa published by \citet{rog11}. This run places the
forming planet at 4 AU with a solid surface density of 6 g cm$^{-2}$ and
a corresponding $M_\mathrm{iso} = 2.4 $  M$_\oplus$ (these parameters are
the same as in the current Runs 4 and 4H). The physics included
in the calculation is essentially the same as for the runs shown in 
Table 1. The Run IIa was cut off
at 2 Myr with  $M_\mathrm{core} =2.65$ M$_\oplus$ 
and  $M_\mathrm
{env} = 0.54$ M$_\oplus$.  The continuation to 3 Myr shows    
 that $\dot M_\mathrm{env}$ remains at a fairly constant value of
$3.5 \times 10^{-7}$ M$_\oplus$ yr$^{-1}$. At 3 Myr the core and
envelope masses are, respectively 2.8 and 0.91 M$_\oplus$. 
Further evolution for another Myr would result in an estimated
$M_\mathrm{env} \approx 1.2$  M$_\oplus$. Thus it is reasonable to
be able to form a planet with the pre-mass-loss characteristics of Kepler-11 f
within a standard disk lifetime. The relatively small core mass
results in a slow envelope accretion rate, so the object is
unlikely to reach the stage of rapid gas accretion.  
Although \citet{lop12} did not calculate the formation phase, this
result is
consistent with their suggestion that formation of the Kepler-11 system
occurred outside the snow line and that the cores contain a
substantial ice component.
 
A rather different result on mass loss is obtained by \citet{chi13}.
Using essentially the same energy-limited mass loss formula as in the
previously-cited papers, but without a full calculation of the thermal
evolution, they find that the amount of mass lost over the
lifetime is given by (their equation 31)
\begin{equation}
\Delta M_\mathrm{env} \sim 0.01 M_\oplus\left(\frac{\epsilon}{0.1}\right)
\left(\frac{R}{5 R_\oplus}\right)^3\left(\frac{10 M_\oplus}{M}\right)
\left(\frac{0.2 \textrm{AU}}{a_p}\right)^2
\end{equation}
where $R$ is the radius during the time before 0.1 Myr and $M$ is the
total planet mass. Taking $M=2$ M$_\oplus$, $\epsilon=0.1$, $a_p=0.25$ AU, 
 and estimating $R= 4$ 
R$_\oplus$ from Figure \ref{fig:2} with $M_\mathrm{env} \approx 0.08$
 M$_\oplus$, we obtain $\Delta M_\mathrm{env} \approx 0.016$  M$_\oplus$.
Note that all of these mass loss estimates involve considerable
uncertainty. 
 
In summary, the mass loss estimates from \citet{iko12} and \citet{chi13}
imply that Kepler-11 f initially had $M_\mathrm{env} \approx 0.1$
 M$_\oplus$ and could easily have formed within 2 AU as an object with
a rock/iron core and a low-mass H/He envelope; note that our calculation at 2 AU 
gives an envelope mass just after formation of 0.16 M$_\oplus$. On the other hand,
with the mass loss estimate from \citet{lop12}, or in any case with
mass loss substantially greater than 0.1  M$_\oplus$, the object  is
much more likely to have
formed exterior to the snow line with an ice-rich core.
In both cases,  migration to the present orbital location is indicated.
{\it In situ} formation is possible only under a very  limited set of
assumptions, including (1) the planet's mass is near the one-standard-deviation 
observational upper limit (2.8 M$_\oplus$), (2) mass loss after formation is
negligible, and (3) either a) the solid surface density in the disk at the
formation location is about 1700 g cm$^{-2}$, which is required to
build a 2.8 M$_\oplus$ core   at 0.25 AU (Eq. \ref{eq:iso}),  or  b) an equivalent
mass in solids was delivered to the planet via gas drag.

\section{Summary and Conclusions}

We investigate the formation and evolution, up to 4 Gyr,
of planets with core masses of 2.2--2.5 M$_\oplus$ and
with core compositions of either iron and rock or iron and
rock and ice. Gas accretion onto the cores is calculated
with a detailed envelope model that includes the 
effects of dust settling and coagulation in the
opacity calculation. The accretion is carried out at
distances of 4, 2, 1, and 0.5 AU from the central
star.  At a fixed cutoff time for the protoplanetary
disk (2 Myr), the amount of accreted gas ranges from
0.037 M$_\oplus$ at 0.5 AU  to 0.16 M$_\oplus$ at 2 AU. Previous
results \citep{rog11} at 4 AU give an envelope mass of 0.54
M$_\oplus$ at the same time with a slightly higher core mass.
For $M_\mathrm{env}$ in the range 0.037 to 0.16 M$_\oplus$, 
final radii, after 4 Gyr, fall in the range 2--6 
R$_\oplus$, depending on envelope mass and core composition,
 as well as on the 
assumed value of the equilibrium surface temperature
during the constant-mass evolution phase. These radii
are in general in agreement with those observed
by {\it Kepler} for sub-Neptune-type planets. 

The values of mass that we calculate
fall within the range of observed values for
Kepler-11 f, with mass 2.0 (+0.8, --0.9) M$_\oplus$,
radius 2.48 (+0.02, --0.03) R$_\oplus$. This planet
orbits at 0.25 AU with an equilibrium temperature of
525 K. The envelope mass is estimated to be about
0.08 M$_\oplus$ \citep{lop12}. Our models indicate that if the planet had
formed {\it in situ},
it could not have accreted enough envelope
mass to account for its present radius, even  if XUV-driven mass 
loss were not important. If the actual
planet mass were near the one-standard-deviation upper limit
(2.8 M$_\oplus$) then {\it in situ} formation (without
mass loss) could have occurred.  On the other hand,
some models indicate that  substantial mass loss
 ($\sim 1$ M$_\oplus$) from the 
envelope would have occurred at its present orbital
position.  In that case, {\it in situ} formation is not
possible, and the planet probably formed beyond the snow line 
at $\approx 4$ AU with a rock/iron/ice core, and, during formation, 
migrated inward. If the mass loss at the current orbit were 
only moderate ($< 0.1 $  M$_\oplus$), then the planet
could have formed between 1 and 2 AU with a rock/iron core, 
coupled with migration inward. Furthermore, if the planet
formed at 4 AU according to the core-nucleated accretion model,
the required solid surface density would have been about     
twice that in the minimum-mass solar nebula. If it had formed
{\it in situ}, that density would have to be $\approx 25$ times
that in the MMSN, corresponding to 4 times that in the minimum-mass
extrasolar nebula of \citet{chi13}. Thus the results of this paper support a
migration history for the planet Kepler-11 f.

\acknowledgments

Primary funding for this project was provided by the NASA Origins of 
Solar Systems Program grant  NNX11AK54G (P. B., J. L.).
 P. B. acknowledges additional support
from NSF grant AST0908807.

\end{document}